\documentclass[11pt]{article}
\usepackage{amssymb,latexsym,amsmath,amsbsy}
\usepackage[dvips]{graphicx}
\usepackage{cite}
\newtheorem{theo}{Theorem}

\newtheorem{coro}[theo]{Corollary}

\newtheorem{prop}[theo]{Proposition}
\def\nn{\nonumber}
\def\deg{\mathop{\rm deg}\nolimits}

\def\ch{\mathop{\rm char}\nolimits}

\def\qdots{\mathinner{\mkern1mu\raise1pt\vbox{\kern7pt\hbox{.}}\mkern2mu
 \raise4pt\hbox{.}\mkern2mu\raise7pt\hbox{.}\mkern1mu}}
\def\Z{{\mathbb Z}}
\def\N{{\mathbb N}}

\def\C{{\mathbb C}}
\def\gl{\mathfrak{gl}}

\def\u{\mathfrak{u}}

\def\so{\mathfrak{so}}
\def\sp{\mathfrak{sp}}
\def\osp{\mathfrak{osp}}
\def\lb{[\![}
\def\rb{]\!]}

\begin{document}
\begin{center}
{\Large \bf
The parastatistics Fock space and explicit Lie superalgebra representations} \\[5mm]
{\bf N.I.~Stoilova}
\\[1mm]
Institute for Nuclear Research and Nuclear Energy, \\ Boul. Tsarigradsko Chaussee 72,
1784 Sofia, Bulgaria\\
E-mail: stoilova@inrne.bas.bg
\end{center}


\begin{abstract}
It is known that the defining triple relations  of $m$ pairs of parafermion operators $f_i^\pm$ 
and $n$ pairs of paraboson operators $b_j^\pm$ with relative parafermion relations 
can be considered as  defining relations for the Lie superalgebra
$\osp(2m+1|2n)$ in terms of $2(m+n)$ generators. With the common Hermiticity conditions, this means
that the parastatistics Fock space of order $p$ corresponds to an 
infinite-dimensional unitary irreducible representation
$V(p)$ of $\osp(2m+1|2n)$, with lowest weight $(-\frac{p}{2},\ldots,- \frac{p}{2}|\frac{p}{2},\ldots,\frac{p}{2})$.
These representations (also in the simplest case $m=n=1$)
had never been constructed due to computational difficulties, despite their importance.
In the present paper we solve partially the problem in the general case 
using group theoretical techniques,
in which the $\u(m|n)$ subalgebra of $\osp(2m+1|2n)$ plays a crucial role: a set of Gelfand-Zetlin
patterns of $\u(m|n)$ can be used to label the basis vectors of $V(p)$.
An explicit and elegant construction of these representations $V(p)$ for $m=n=1$,
and the actions or matrix elements of the $\osp(3|2)$ generators are given.
\end{abstract}

\vskip 10mm
\noindent Running title: Parastatistics and Lie superalgebras


\setcounter{equation}{0}
\section{Introduction} \label{sec:Introduction}%

In 1953
Green~\cite{Green} generalized the ordinary Fermi-Dirac and Bose-Einstein
statistics introducing the parafermion and paraboson statistics. 
Parastatistics were also formulated algebraically in terms of generators 
and relations.
The operators of parafermion statistics $f_j^\pm$, $j=1,2,\ldots, m$, satisfying
\begin{equation}
[[f_{ j}^{\xi}, f_{ k}^{\eta}], f_{l}^{\epsilon}]=\frac 1 2
(\epsilon -\eta)^2
\delta_{kl} f_{j}^{\xi} -\frac 1 2  (\epsilon -\xi)^2
\delta_{jl}f_{k}^{\eta},  
\label{f-rels}
\end{equation}
where $j,k,l\in \{1,2,\ldots,m\}$ and $\eta, \epsilon, \xi \in\{+,-\}$ (to be interpreted as $+1$ and $-1$
in the algebraic expressions $\epsilon -\xi$ and $\epsilon -\eta$), are generating elements of
the orthogonal  Lie algebra $\mathfrak{so}(2m+1)$~\cite{Kamefuchi,Ryan}. In a similar way 
$n$ paraboson operators $b_j^\pm$,   satisfying
\begin{equation}
[\{ b_{ j}^{\xi}, b_{ k}^{\eta}\} , b_{l}^{\epsilon}]= (\epsilon -\xi) \delta_{jl} b_{k}^{\eta} 
 +  (\epsilon -\eta) \delta_{kl}b_{j}^{\xi}, 
\label{b-rels}
\end{equation}
are generating elements of the orthosymplectic Lie superalgebra 
$\mathfrak{osp}(1|2n)$~\cite{Ganchev}. 
The important objects to  construct are the generalizations of the fermion and boson Fock spaces. 
Parafermion and paraboson Fock spaces are characterized by a parameter $p$, the order of the parastatistics. 
Although already in 1953 Green proposed a general approach, known nowadays as Green ansatz~\cite{Green}, for the construction 
of the parafermion and paraboson Fock spaces their structure was not known until a few years ago. The 
difficulties of the Green ansatz are connected to the problem of finding proper bases of irreducible constituents of 
$p$-fold tensor products~\cite{Greenberg, KD}, and did not lead to a solution of the problem.
Recently, 
for the case of parafermions, this explicit construction of the Fock space of order $p$ was 
given in~\cite{parafermion}, and for  parabosons in~\cite{paraboson}.
The solutions use the algebraic formulations of parafermion and paraboson statistics. 
As a concequence, the parafermion Fock space of order $p$ is the finite-dimensional unitary irreducible 
representation (unirrep) of $\so(2m+1)$ with lowest weight $(-\frac{p}{2},-\frac{p}{2},\ldots, -\frac{p}{2})$, and the 
paraboson Fock space is the infinite-dimensional unirrep of $\osp(1|2n)$ also with lowest weight $(\frac{p}{2},\frac{p}{2},\ldots, \frac{p}{2})$.
The constructions use the branchings $\so(2m+1)\supset  \u(m)$ for the parafermion  Fock space and $\osp(1|2n)\supset \sp(2n)\supset \u(n)$ for the paraboson Fock space.
The results are complete descriptions of proper bases and the explicit action of the parafermion and paraboson operators in the
corresponding basis~\cite{parafermion, paraboson}. 

As a next step, it  is natural to extend these results to a system consisting of parafermions $f_j^\pm$ and
parabosons $b_j^\pm$. The commutation relations among paraoperators were studied by Greenberg and Messiah~\cite{GM}.
As a concequence of some natural assumptions they came to the result that for each pair of paraoperators there can exist at most four types of relative  commutation relations: straight commutation, straight anticommutation, relative paraboson, and relative parafermion relations. 
The case with relative paraboson relations and the corresponding Fock representations has been investigated in~\cite{YJ}-\cite{KK}. In the present paper we consider relative parafermion relations.
It was proved by Palev~\cite{Palev1} that  $m$  
parafermions $f_j^\pm\equiv c_j^\pm$~(\ref{f-rels}) and $n$ parabosons $b_j^\pm\equiv c_{m+j}^\pm$~(\ref{b-rels}) 
with relative parafermion relations
lead to the result that they
 generate the
orthosymplectic Lie superalgebra $\mathfrak{osp}(2m+1|2n)$. Then the parastatistics Fock space of order $p$ corresponds to 
an infinite-dimensional unitary representation of $\mathfrak{osp}(2m+1|2n)$ and it can be constructed explicitly 
using similar techniques as in~\cite{paraboson, parafermion}, namely using the branching $\mathfrak{osp}(2m+1|2n)\supset \mathfrak{gl}(m|n)$,
an induced representation construction,
a basis description for the covariant tensor representations of  $\mathfrak{gl}(m|n)$,
Clebsch-Gordan coefficients of $\mathfrak{gl}(m|n)$, and  the method of reduced matrix elements. 
The covariant tensor representations 
of the Lie superalgebra $\mathfrak{gl}(m|n)$ in an explicit form 
and the relevant Clebsch-Gordan coefficients (namely those corresponding to the tensor product $V([\mu]^{r})\otimes V([1,0,\ldots,0])$, where $V([\mu]^{r})$ is any  $\mathfrak{gl}(m|n)$
irreducible covariant tensor representation and $V([1,0,\ldots,0])$ is the representation of 
$\mathfrak{gl}(m|n)$ with highest weight $(1,0,\ldots,0)$) 
were constructed and found in~\cite{CGC}. 

The structure of the paper is the  following. In section~2, we define 
the parastatistics Fock space $V(p)$. In section~3, we consider  the important
relation between parastatistics operators and the Lie superalgebra $\osp(2m+1|2n)$, and give a 
description of $V(p)$ in terms of representations of $\osp(2m+1|2n)$. 
Section~4 is devoted to the analysis of the
representations $V(p)$ for $\osp(2m+1|2n)$ and to finding the matrix elements for $m=n=1$, where 
the main computational result is given in Theorem~\ref{prop-main}. We conclude the paper with some final remarks.

\setcounter{equation}{0}
\section{The parastatistics Fock space $V(p)$} \label{sec:Fock}

Before introducing the parastatistics Fock space, we will consider the Fock space $V(1)$
corresponding to a system of 
$m$ pairs Fermi operators $F_i^\pm, i=1,2,\ldots,m$ $(\{ a, b\}=ab+ba)$
\begin{equation}
\{ F_i^-, F_k^+ \} =\delta_{ik},\qquad \{ F_i^-, F_k^- \} =\{ F_i^+, F_k^+ \} =0
\label{FF}
\end{equation}
and $n$ pairs Bose operators $B_j^\pm, j=1,2,\ldots,n$ $([ a, b ]=ab-ba)$
\begin{equation}
[ B_j^-, B_l^+ ] =\delta_{jl},\qquad [ B_j^-, B_l^- ] =[ B_j^+, B_l^+ ] =0,
\label{BB}
\end{equation}
which
mutually anticommute 
\begin{equation}
 \{ F_i^\xi,B_j^\eta\}=0, \;\;\xi, \eta =\pm; \;\; i=1,\ldots,m; \;\; j=1,\ldots,n.
 \label{mn-fermionboson}
\end{equation}
The Fock space $V(1)$ is  defined as a Hilbert space with vacuum vector $|0\rangle$, 
with
\begin{equation}
\langle 0|0\rangle=1, \qquad F_i^- |0\rangle = B_j^- |0\rangle=0, \quad (F_i^\pm)^\dagger = F_i^\mp, \quad (B_j^\pm)^\dagger = B_j^\mp.
\end{equation}
The Hilbert space is irreducible under the action of the algebra
spanned by the elements $1, F_i^\pm, B_j^\pm$.
 \;A set of (orthogonal and normalized) basis vectors of this space is given by
\begin{equation}
 |\theta_1,\ldots,\theta_m,k_1,\ldots,k_n \rangle = \frac{(F_1^+)^{\theta_1}\cdots(F_m^+)^{\theta_m} (B_1^+)^{k_1}\cdots(B_n^+)^{k_n}}{\sqrt{k_1!\cdots k_n!}}|0\rangle,
 \quad \theta_i=0,1;\;  k_j \in\Z_+.
 \label{basis-mn}
\end{equation}
A straightforward calculation gives
\begin{align}
&F_i^+|\ldots,\theta_i,\ldots,k_j,\ldots \rangle = (-1)^{\theta_1+\ldots+\theta_{i-1}}\sqrt{1-\theta_i}
|\ldots,\theta_i+1,\ldots,k_j,\ldots \rangle , \\
&F_i^-|\ldots,\theta_i,\ldots,k_j,\ldots \rangle = (-1)^{\theta_1+\ldots+\theta_{i-1}}\sqrt{\theta_i}
|\ldots,\theta_i-1,\ldots,k_j,\ldots \rangle , \\
&B_j^+|\ldots,\theta_i,\ldots,k_j,\ldots \rangle = (-1)^{\theta_1+\ldots+\theta_{m}}\sqrt{1+k_j}
|\ldots,\theta_i,\ldots,k_j+1,\ldots \rangle , \\
&B_j^-|\ldots,\theta_i,\ldots,k_j,\ldots \rangle = (-1)^{\theta_1+\ldots+\theta_{m}}\sqrt{k_j}
|\ldots,\theta_i,\ldots,k_j-1,\ldots \rangle \label{trans}.
\end{align} 

This Fock space is a certain unirrep of the Lie superalgebra
$\osp(2m+1|2n)$~\cite{Palev1}, with lowest weight $(-\frac{1}{2},\ldots,-\frac{1}{2}|\frac{1}{2},\ldots,\frac{1}{2})$.

We are interested in a system of $m$ pairs of parafermion  operators 
$f_i^\pm\equiv c_i^\pm$ $, i=1,\ldots,m$ and $n$ pairs of paraboson operators $b_j^\pm\equiv c_{m+j}^\pm$ $, j=1,\ldots,n$
with relative parafermion relations among them. The defining triple relations for such a system
are given by~\cite{Palev1}
\begin{eqnarray}
&& \lb\lb c_{ j}^{\xi}, c_{ k}^{\eta}\rb , c_{l}^{\epsilon}\rb =-2
\delta_{jl}\delta_{\epsilon, -\xi}\epsilon^{\langle l \rangle} 
(-1)^{\langle k \rangle \langle l \rangle }
c_{k}^{\eta} +2  \epsilon^{\langle l \rangle }
\delta_{kl}\delta_{\epsilon, -\eta}
c_{j}^{\xi},  \label{para}\\
&& \qquad\qquad \xi, \eta, \epsilon =\pm\hbox{ or }\pm 1;\quad j,k,l=1,\ldots,n+m, \nn 
\end{eqnarray} 
where
\begin{equation}
\lb a, b \rb = ab-(-1)^{{\deg(a)
\deg(b)}}ba \;\;  \rm{and } \;\; \deg(c_i^\pm)\equiv \langle i\rangle= \left\{ \begin{array}{lll}
 {0} & \hbox{if} & j=1,\ldots ,m \\ 
 {1} & \hbox{if} & j=m+1,\ldots ,n+m.
 \end{array}\right.
\end{equation}
In the case $j,k,l=1,\ldots,m$~(\ref{para}) reduces to~(\ref{f-rels}) and in the case $j,k,l=m+1,\ldots,m+n$~(\ref{para}) reduces to~(\ref{b-rels}). 

The parastatistics Fock space $V(p)$ is the Hilbert space with vacuum vector $|0\rangle$, 
defined by means of ($j,k=1,2,\ldots,m+n$)
\begin{align}
& \langle 0|0\rangle=1, \qquad c_j^- |0\rangle = 0, \qquad (c_j^\pm)^\dagger = c_j^\mp,\nn\\
& \lb c_j^-,c_k^+ \rb |0\rangle = p\delta_{jk}\, |0\rangle, \label{Fock}
\end{align}
and by irreducibility under the action of the algebra spanned by
the elements $c_j^+$, $c_j^-$, $j=1,\ldots,m+n$, subject
to~\eqref{para}. The parameter $p$ is referred to as the order of the parastatistics system
 and for $p=1$ the parastatistics Fock
space $V(p)$ coincides with the Fock space $V(1)$~(\ref{mn-fermionboson})-(\ref{trans}) of $n$ bosons and $m$ fermions with unusual grading 
which anticommute.

Constructing a basis for the parastatistics Fock space $V(p)$ 
for general (integer) $p$-values,
turns out to be a difficult problem,
unsolved so far not only in general bul also for a single parafermion and paraboson.
Even the simpler question of finding the structure of $V(p)$ (weight structure)
is not solved. 
In the present paper we shall solve partially the last problem  for any $m$  pairs 
of parafermions and $n$ pairs of parabosons  and we shall  construct an orthogonal (normalized) basis for $V(p)$,
and give the actions of the generators $c_j^\pm$ on the basis vectors for $m=n=1$.

\section{The Lie superalgebras $B(m|n)$}
\setcounter{equation}{0} \label{sec:B}

The Lie superalgebra 
$B(m|n)\equiv \osp(2m+1|2n)$~\cite{Kac}  consists of matrices of the form
\begin{equation}
\left(\begin{array}{ccccc} a&b&u&x&x_1  \\
c&-a^t&v&y&y_1\\
-v^t&-u^t&0&z&z_1\\
y_1^t&x_1^t&z_1^t&d&e\\
-y^t&-x^t&-z^t&f&-d^t
\end{array}\right),
\label{osp}
\end{equation}
where $a$ is any $(m\times m)$-matrix, $b$ and $c$ are antisymmetric $(m\times m)$-matrices, 
$u$ and $v$ are $(m\times 1)$-matrices, $x,y,x_1,y_1$ are $(m\times n)$-matrices, $z$ and
$z_1$ are $(1\times n)$-matrices, $d$ is any $(n\times n)$-matrix,
and $e$ and $f$ are symmetric  $(n\times n)$-matrices. The even elements have 
$x=y=x_1=y_1=0$, $z=z_1=0$ and the odd elements are those with 
$a=b=c=0$, $u=v=0$, $d=e=f=0$. Denote the row and column indices running from $1$ to $2m+2n+1$ 
 and  by $e_{ij}$ the matrix with zeros everywhere except
a $1$ on position $(i,j)$.
The Cartan subalgebra $H$ of $\osp(2m+1|2n)$ is the
subspace of diagonal matrices  with basis $h_i=e_{ii}-e_{i+m,i+m}, \; i=1,\ldots,m;$ 
$h_{m+i}=e_{2m+1+j,2m+1+j}-e_{2m+1+n+j,2m+1+n+j}, \;j=1,\ldots,n.$ 
In terms of the dual basis 
 $\epsilon_i,  i=1,\ldots, m; $ 
$\delta_j, j=1,\ldots,n$ of $H^*$, the even 
root vectors and corresponding roots of $osp(2m+1|2n)$ are given by
\begin{eqnarray*}
e_{jk}-e_{k+m,j+m} & \leftrightarrow &\epsilon_j -\epsilon_k, \qquad j\neq k=1,\ldots ,m,\\
e_{j,k+m}-e_{k,j+m} & \leftrightarrow &\epsilon_j +\epsilon_k, \qquad j<k=1,\ldots ,m,\\
e_{j+m,k}-e_{k+m,j} & \leftrightarrow &-\epsilon_j -\epsilon_k, \qquad j<k=1,\ldots ,m,\\
e_{j,2m+1}-e_{2m+1,j+m} & \leftrightarrow &\epsilon_j,  \qquad j=1,\ldots ,m,\\
e_{j+m,2m+1}-e_{2m+1,j} & \leftrightarrow & -\epsilon_j , \qquad j=1,\ldots ,m,\\
e_{2m+1+j,2m+1+k}-e_{n+2m+1+k,n+2m+1+j} & \leftrightarrow &\delta_j -\delta_k, 
\qquad j\neq k=1,\ldots ,n,\\
e_{2m+1+j,2m+1+k+n}+e_{2m+1+k,2m+1+j+n} & \leftrightarrow 
&\delta_j +\delta_k, \qquad j\leq k=1,\ldots ,n,\\
e_{2m+1+n+j,2m+1+k}+e_{2m+1+n+k,2m+1+j} & \leftrightarrow &-
\delta_j -\delta_k, \qquad j\leq k=1,\ldots ,n,
\end{eqnarray*}
and the odd ones by
\begin{eqnarray*}
e_{j,2m+1+k}-e_{2m+1+n+k,j+m} & \leftrightarrow &\epsilon_j -\delta_k, 
\qquad j=1,\ldots ,m; \ k=1,\ldots ,n,\\
e_{m+j,2m+1+k}-e_{2m+1+n+k,j} & \leftrightarrow &-\epsilon_j -\delta_k, 
\qquad j=1,\ldots ,m; \ k=1,\ldots , n,\\
e_{2m+1,2m+1+k}-e_{2m+1+n+k,2m+1} 
& \leftrightarrow &-\delta_k, \qquad k=1,\ldots ,n,\\
e_{j,2m+1+n+k}+e_{2m+1+k,m+j} & \leftrightarrow &\epsilon_j +\delta_k, 
\qquad j=1,\ldots ,m; \ k=1,\ldots, n,\\
e_{m+j,2m+1+n+k}+e_{2m+1+k,j} & \leftrightarrow &-\epsilon_j +\delta_k, 
\qquad j=1,\ldots ,m; \ k=1,\ldots, n,\\
e_{2m+1,2m+1+n+k}+e_{2m+1+k,2m+1} & \leftrightarrow &\delta_k, \qquad k=1,\ldots ,n.
\end{eqnarray*}
If we introduce the following multiples of the  even vectors with roots 
$\pm \epsilon_j, \; j=1,\ldots, m $ 
\begin{align}
&c_{j}^+=f_{j}^+= \sqrt{2}(e_{j, 2m+1}-e_{2m+1,j+m}), \nn\\
&c_{j}^-=f_{j}^-= \sqrt{2}(e_{2m+1,j}-e_{j+m,2m+1}), \;
\label{f-as-e}
\end{align}
and of the  odd vectors with roots 
$\pm \delta_j, j=1,\ldots,n$
\begin{align}
&
c_{m+j}^+=b_{j}^+= \sqrt{2}(e_{2m+1,2m+1+n+j}+e_{2m+1+j,2m+1}), \nn\\
&c_{m+j}^-=b_{j}^-= \sqrt{2}(e_{2m+1,2m+1+ j}-e_{2m+1+n+j,2m+1}), \; 
\label{b-as-e}
\end{align}
it is easy to verify that these operators satisfy the  triple relations~(\ref{para}).

Moreover, the following holds~\cite{Palev1}

\begin{theo}[Palev]
As a Lie superalgebra defined by generators and relations, 
$\osp(2m+1|2n)$ is generated by $2m+2n$ elements $c_j^\pm$  subject to the parastatistics relations
\begin{eqnarray}
&& \lb\lb c_{ j}^{\xi}, c_{ k}^{\eta}\rb , c_{l}^{\epsilon}\rb =-2
\delta_{jl}\delta_{\epsilon, -\xi}\epsilon^{\langle l \rangle} 
(-1)^{\langle k \rangle \langle l \rangle }
c_{k}^{\eta} +2  \epsilon^{\langle l \rangle }
\delta_{kl}\delta_{\epsilon, -\eta}
c_{j}^{\xi},  \label{paraosp}\\
&& \qquad\qquad \xi, \eta, \epsilon =\pm\hbox{ or }\pm 1;\quad j,k,l=1,\ldots,n+m. \nn 
\end{eqnarray} 
\end{theo}
The paraoperators $c_j^+$ are  positive  root vectors, and the
$c_j^-$ are  negative  root vectors.

We are  interested in the construction of 
the parastatistics  Fock space $V(p)$ defined by~(\ref{Fock}):
\begin{align}
& \langle 0|0\rangle=1, \qquad c_j^- |0\rangle = 0, \qquad (c_j^\pm)^\dagger = c_j^\mp,\nn\\
& \lb c_j^-,c_k^+ \rb |0\rangle = p\delta_{jk}\, |0\rangle \nn
\end{align}
and by irreducibility under the action of the algebra spanned by
the elements $c_j^+$, $c_j^-$, $j=1,\ldots,m+n$, subject
to~\eqref{para}.
It is straightforward to see that
\begin{equation}
[c_i^-,c_i^+]=-2 h_i, \; i=1,\ldots,m,\;\; {\rm and} \;\; \{c_{m+j}^-,c_{m+j}^+]=2 h_{m+j},  \; j=1,\ldots,n.
\label{bbh}
\end{equation}
Therefore we have 
\begin{coro}
The parastatistics Fock space $V(p)$ is the unitary irreducible representation of
$\osp(2m+1|2n)$ with lowest weight $(-\frac{p}{2},\ldots, -\frac{p}{2}|\frac{p}{2},\ldots, \frac{p}{2})$.
\end{coro} 

In order to construct the representations $V(p)$ in general one can use an induced module procedure.
The relevant subalgebras of $\osp(2m+1|2n)$ are easy to describe by means of the
the generators $c_j^\pm$. 
\begin{prop}
A basis for the even subalgebra $\so(2m+1)\oplus \sp(2n)$ of $\osp(2m+1|2n)$ is given by the elements
\begin{equation}
[c_i^\xi, c_k^\eta],\; c_l^\epsilon, \; i,k,l=1,\ldots,m;\quad \{c_{m+j}^\xi, c_{m+s}^\eta\},\;j,s=1,\ldots,n,
\; \xi,\eta =\pm.
\label{sp2n}
\end{equation}
The $(m+n)^2$ elements 
\begin{equation}
\lb c_j^+, c_k^-\rb \qquad(j,k=1,\ldots,m+n)
\label{un}
\end{equation} 
are a basis for the  subalgebra $\u(m|n)$.
\end{prop}

Note that with $\frac{1}{2}\lb c_j^+,c_k^-\rb=E_{jk}$, the triple relations~\eqref{paraosp} imply the
relations $\lb E_{ij},E_{kl}\rb =\delta_{jk}E_{il}-(-1)^{deg(E_{ij})deg(E_{kl})}\delta_{li}E_{kj}$. 
Therefore, the 
elements $\lb c_j^+,c_k^-\rb$ form, up to a factor 2, the standard $\u(m|n)$ or $\gl(m|n)$
basis elements.

The superalgebra $\u(m|n)$ is, algebraically, the same as the general linear Lie superalgebra $\gl(m|n)$. 
However the condition $(c_j^\pm)^\dagger=c_j^\mp$ implies that we are dealing here with the ``compact form'' $\u(m|n)$.

Let us extend the subalgebra $\u(m|n)$ to a parabolic subalgebra ${\cal P}$ of $\osp(2m+1|2n)$
\begin{equation}
{\cal P} = \hbox{span} \{ c_j^-, \lb c_j^+, c_k^-\rb,  \lb c_j^-, c_k^-\rb \;|\;
j,k=1,\ldots,m+n \}.
\label{P}
\end{equation}

Since  $\lb c_j^-,c_k^+\rb |0\rangle = p\,\delta_{jk}\, |0\rangle$, with 
$[c_i^-,c_i^+]=-2 h_i,$ \; $i=1,\ldots,m,$\;  and\; $\{c_{m+j}^-,c_{m+j}^+]=2 h_{m+j},$ \;$ j=1,\ldots,n$,
the space spanned by $|0\rangle$ is a trivial one-dimensional $\u(m|n)$ module $\C |0\rangle$
of weight $(-\frac{p}{2},\ldots, -\frac{p}{2}|\frac{p}{2},\ldots, \frac{p}{2})$.
As $c_j^- |0\rangle =0$,  the $\u(m|n)$ module $\C |0\rangle$ can be extended to a one-dimensional ${\cal P}$ module.
The induced $\osp(2m+1|2n)$ module $\overline V(p)$ is  defined by 
\begin{equation}
 \overline V(p) = \hbox{Ind}_{\cal P}^{\osp(2m+1|2n)} \C|0\rangle.
 \label{defInd}
\end{equation}
This is an $\osp(2m+1|2n)$ representation with lowest weight $(-\frac{p}{2},\ldots, -\frac{p}{2}|\frac{p}{2},\ldots, \frac{p}{2})$.
By the Poincar\'e-Birkhoff-Witt theorem~\cite{Kac1}, it is easy to write a basis for  $\overline V(p)$
\begin{align}
& (c_1^+)^{k_1}\cdots (c_{m+n}^+)^{k_{m+n}} (\lb c_1^+,c_2^+\rb)^{k_{12}}  (\lb c_1^+,c_3^+\rb)^{k_{13}} \cdots 
(\lb c_{m+n-1}^+,c_{m+n}^+\rb)^{k_{m+n-1,m+n}} |0\rangle, \label{Vpbasis}\\
& \qquad k_1,\ldots,k_{m+n},k_{12},k_{13}\ldots,k_{m-1,m},k_{m+1,m+2},k_{m+1,m+3}\ldots,k_{m+n-1,m+n}  \in \Z_+, \nn\\
& \qquad k_{1,m+1},k_{1,m+2}\ldots,k_{1,m+n},k_{2,m+1},\ldots,k_{m,m+n}=0,1. \nn
\end{align}
The problem and difficulty  come from the fact that
in general $\overline V(p)$ is not a simple module (i.e.\ not an irreducible representation) of
$\osp(2m+1|2n)$. Let $M(p)$ be the maximal nontrivial submodule of $\overline V(p)$. Then the
simple module (irreducible module), corresponding to the parastatistics Fock space, is
\begin{equation}
V(p) = \overline V(p) / M(p).
\label{Vp}
\end{equation}
The problem is  to determine the vectors belonging to $M(p)$, and therefore to find 
the structure of $V(p)$. 

\setcounter{equation}{0}
\section{Parastatistics Fock space of  $\osp(2m+1|2n)$. Matrix elements for $m=n=1$} \label{sec:osp32}

Consider the induced module $\overline V(p)$ in the case $m=n=1$, with basis vectors
\begin{equation}
|k,l,\theta\rangle \equiv  (c_1^+)^{k}(c_2^+)^{l} ([c_1^+,c_2^+])^{\theta} |0\rangle,
\qquad k,l\in\Z_+; \theta =0,1.
\label{kltheta}
\end{equation}
The weight of this vector is
\begin{equation}
(-\frac{p}{2}| \frac{p}{2}) + k \epsilon_1+ l\delta_1 +\theta (\epsilon_1+\delta_1).
\label{wt-kltheta}
\end{equation}
The {\em level} of such a  vector is defined as $k+l+2\theta$.
 The actions of the generators
$c_1^\pm$, $c_2^\pm$ on the basis vectors $|k,l,\theta\rangle$ can be computed, using the triple relations.
For the positive root vectors the computations are easy
\begin{align}
& c_1^+ |k,l,\theta\rangle = |k+1,l,\theta\rangle,\nn\\
& c_2^+ |k,l,0\rangle = |k,l+1,0\rangle-(-1)^lk|k-1,l,1\rangle     ,\nn\\
& c_2^+ |k,l,1\rangle = |k,l+1,1\rangle.
\label{c+kltheta}
\end{align}
However for the negative root vectors this requires some tough computations
\begin{align}
& c_1^- |k,l,0\rangle = k(p-k+1)|k-1,l,0\rangle,\nn\\
& c_1^- |k,l,1\rangle = k(p-k-1)|k-1,l,1\rangle+(-1)^l2|k,l+1,0\rangle     ,\nn\\
& c_2^- |k,2l,0\rangle = -2kl|k-1,2l-2,1\rangle+2l|k,2l-1,0\rangle     ,\nn\\
& c_2^- |k,2l+1,0\rangle = 2kl|k-1,2l-1,1\rangle+(p+2l-2k)|k,2l,0\rangle     ,\label{b-lktheta}\\
& c_2^- |k,2l,1\rangle = 2l|k,2l-1,1\rangle -2|k+1,2l,0\rangle     ,\nn\\
& c_2^- |k,2l+1,1\rangle = (p+2l-2k)|k,2l,1\rangle+2|k+1,2l+1,0\rangle.\nn
\end{align}

Using $\langle0|0\rangle=1$ and $(c_i^\pm)^\dagger = c_i^\mp$ we can compute
``inner products'' of the vectors $|k,l,\theta\rangle$ 
\begin{align}
& \langle k,l,\theta|k, l,\theta \rangle \equiv \big( (c_1^+)^k (c_2^+)^l ([c_1^+,c_2^+ ])^\theta|0\rangle,
(c_1^+)^k (c_2^+)^l ([c_1^+,c_2^+ ])^\theta|0\rangle \big)\nn\\
& = \big( ([c_2^-,c_1^- ])^\theta(c_2^-)^l (c_1^-)^k(c_1^+)^k (c_2^+)^l ([c_1^+,c_2^+ ])^\theta|0\rangle, |0\rangle .
\end{align}
Straightforward long computations give:
\begin{align}
& \langle k,2l,0|k, 2l,0\rangle = k!(p-k+1)_k 2^{2l}l!(\frac{p}{2})_l,\label{k2l0}\\
& \langle k,2l+1,0|k, 2l+1,0\rangle = k!(p-k+1)_k 2^{2l+1}l!(\frac{p}{2})_{l+1},\label{k2l+10}\\
& \langle k,2l,1|k, 2l,1\rangle = 4k!(p-k)_{k+1} 2^{2l}l!(\frac{p}{2}+1)_l,\label{k2l1}\\
& \langle k,2l+1,1|k, 2l+1,1\rangle = 4k!(p-k)_{k+1} 2^{2l}l!(p-k+2l+1)(\frac{p}{2}+1)_{l},\label{k2l+11}
\end{align}
where the symbol $(a)_k = a(a+1)\cdots (a+k-1)$ 
is the common Pochhammer symbol.

{}From~(\ref{k2l0}) ($l=0, k=1$) it follows that $p$ should be a positive number, otherwise the inner product is not positive definite.
In a similar way whenever $p$ is fixed then Eqs. (\ref{k2l0})-(\ref{k2l+11})) give that $k=0,1,\ldots,p-\theta$.
Now having in mind that 
vectors of different weight have inner product zero  we must find the inner product of vectors with one and the
same weight.  
At level $0$ there is one vector of weight $(-\frac{p}{2}| \frac{p}{2})$ only, $|0,0,0\rangle = |0\rangle$;
at level 1 there is one vector of weight $(-\frac{p}{2}+1| \frac{p}{2})$ and one of weight
$(-\frac{p}{2}| \frac{p}{2}+1)$; 
at level 2 there is one vector of weight $(-\frac{p}{2}+2| \frac{p}{2})$, one of weight
$(-\frac{p}{2}| \frac{p}{2}+2)$, but two vectors of weight $(-\frac{p}{2}+1| \frac{p}{2}+1)$.
The inner products of the latter are given by:
\begin{align}
& \langle 1,1,0|1,1,0\rangle=p^2,\ \langle 1,1,0|0,0,1\rangle=2p,\ \langle 0,0,1|0,0,1\rangle=4p.
\label{level2}
\end{align}
The matrix of inner products of the vectors of weight $(-\frac{p}{2}+1| \frac{p}{2}+1)$
has determinant
\begin{equation}
\det \left(\begin{array}{cc} p^2 & 2p \\ 2p & 4p \end{array}\right) = 4p^2(p-1).
\label{detlevel2}
\end{equation}
This matrix is positive definite only if $p>1$. 
Therefore, for $p>1$ both vectors of weight $(-\frac{p}{2}+1| \frac{p}{2}+1)$ belong to $V(p)$;
but for $p=1$ one vector ($2|1,1,0\rangle - |0,0,1\rangle$) belongs to $M(p)$ and 
the subspace of $V(p)$ of weight $(-\frac{p}{2}+1| \frac{p}{2}+1)$ is one-dimensional.

We can continue this analysis level by level, but the computations become 
complicated and in order to find a technique that works for arbitrary $m$ and $n$ one should
find a better way of analysing $\overline V(p)$.
For this purpose, we will construct a different basis for $\overline V(p)$.
It is indicated by the character of $\overline V(p)$: this is a formal
infinite series of terms $\nu x_1^{j_1}x_2^{j_2}\ldots x_m^{j_m} y_1^{j_{m+1}}y_2^{j_{m+2}}\ldots y_n^{j_{m+n}}$, 
with $(j_1,\ldots,j_m|j_{m+1},\ldots,j_{m+n})$ a weight of $\overline V(p)$
and $\nu$ the dimension of this weight space.
So the vacuum vector $|0\rangle$ of $\overline V(p)$, of weight $(-\frac{p}{2},\ldots,-\frac{p}{2}|\frac{p}{2},\ldots, \frac{p}{2})$, yields a 
term $x_1^{-\frac{p}{2}}\ldots x_m^{-\frac{p}{2}} y_1^{\frac{p}{2}}\ldots  y_n^{\frac{p}{2}}$
in the character $\ch \overline V(p)$.
Since the basis vectors are given by

\noindent 
$(c_1^+)^{k_1}\cdots (c_{m+n}^+)^{k_{m+n}} (\lb c_1^+,c_2^+\rb)^{k_{12}}  (\lb c_1^+,c_3^+\rb)^{k_{13}} \cdots 
(\lb c_{m+n-1}^+,c_{m+n}^+\rb)^{k_{m+n-1,m+n}} |0\rangle$,

\noindent
where 
$k_1,\ldots,k_{m+n},k_{12},k_{13}\ldots,k_{m-1,m},k_{m+1,m+2},k_{m+1,m+3}\ldots,k_{m+n-1,m+n}  \in \Z_+, \;$

\noindent
$ k_{1,m+1},k_{1,m+2}\ldots,k_{1,m+n},k_{2,m+1},\ldots,k_{m,m+n}=0,1$,
it follows that
\begin{equation}
 \ch \overline V(p) = \frac{(x_1)^{-p/2}\cdots (x_m)^{-p/2}(y_1)^{p/2}\cdots (y_n)^{p/2} \prod_{i,j}(1+x_iy_j)}
{\prod_{i}(1-x_i)\prod_{i<k}(1-x_ix_k)\prod_{j}(1-y_j)\prod_{j<l}(1-y_jy_l)} .
 \label{char-osp32}
\end{equation}
Such expressions have an interesting expansion in terms of supersymmetric Schur functions,
valid for general $m$ and $n$.
\begin{prop}{\bf(C.J. Cummins, R.C. King)}
Let $({\mathbf x})=(x_1, x_2,\ldots, x_m)$ 
and $({\mathbf y})=(y_1, y_2,\ldots, y_n)$ 
 be  sets of $m$ and $n$ variables, respectively. Then~\cite{King}
\begin{equation}
\frac{\prod_{i,j}(1+x_iy_j)}{\prod_{i}(1-x_i)\prod_{i<k}(1-x_ix_k)\prod_{j}(1-y_j)\prod_{j<l}(1-y_jy_l)} = 
\sum_{\lambda \in {\cal{H}}} s_\lambda (x_1,\ldots,x_m|y_1,\ldots,y_n)
= \sum_{\lambda \in {\cal{H}}} s_\lambda ({\mathbf x}|{\mathbf y}). 
\label{Schur}
\end{equation}
In the right hand side, the sum is over all partitions $\lambda$, 
satisfying the so called hook condition $\lambda_{m+1}\leq n$ ($\lambda \in {\cal{H}}$) and 
$s_\lambda({\mathbf x}|{\mathbf y})$  is the supersymmetric Schur function~\cite{Mac} defined by
\[
s_\lambda({\mathbf x| \mathbf y})= \sum_{\tau} s_{\lambda/\tau}({\mathbf x})s_{\tau'}({\mathbf y})
= \sum_{\sigma, \tau} c_{\sigma \tau}^\lambda s_{\sigma}({\mathbf x})s_{\tau'}({\mathbf y}),
\] 
with $l(\sigma)\leq m$; \; $l(\tau')\leq n$; \;$\tau'$  the conjugate partition  to $\tau$;\; $c_{\sigma \tau}^\lambda$  the famous Littlewood-Richardson coefficients;
$|\lambda|= |\sigma|+|\tau|$ and  $s_{\nu}({\mathbf x})$ 
the ordinary Schur function. 
\end{prop}  
The characters of the irreducible covariant $\u(m|n)$ tensor representations $V([\Lambda^\lambda])$, 
which are necessarily finite-dimensional,  
are given by such supersymmetric Schur functions $s_\lambda(x|y), \;\lambda \in {\cal H}$.
The relation between the partitions $\lambda=(\lambda_1, \lambda_2,\ldots), \; \lambda_{m+1}\leq n$
and the highest weights $\Lambda^\lambda$ $\equiv [\mu ]^r\equiv [\mu_{1r},\ldots,
\mu_{mr}|\mu_{m+1,r}\ldots ,\mu_{rr}]$; $r=m+n$, of the irreducible covariant $\u(m|n)$ tensor representations
 is given by~\cite{JHKR}: 
\begin{align}
& \mu_{ir}=\lambda_i, \quad 1\leq i\leq m, \label{hwpart1}\\
& \mu_{m+i,r}=\max\{0, \lambda'_i-m\}, \quad 1\leq i\leq n, \label{hwpart2}
\end{align}
where $\lambda'$ is the partition conjugate~\cite{Mac} to $\lambda$. 
Therefore the expansion~(\ref{Schur}) yields the branching to $\u(m|n)$ of the $\osp(2m+1|2n)$
representation $\overline V(p)$.
This gives a possibility to label the basis vectors of $\overline V(p)$.
For each irreducible covariant $\u(m|n)$ tensor representations one can use the corresponding
Gelfand-Zetlin basis(GZ)~\cite{CGC}. The union of all these GZ basis is then the basis for $\overline V(p)$.
Thus the new basis of $\overline V(p)$ consists of vectors of the form ($p$ is dropped from the notation of the vectors)

\begin{equation}
|\mu)\equiv |\mu)^r = \left|
\begin{array}{lclllcll}
\mu_{1r} & \cdots & \mu_{m-1,r} & \mu_{mr} & \mu_{m+1,r} & \cdots & \mu_{r-1,r}
& \mu_{rr}\\
\mu_{1,r-1} & \cdots & \mu_{m-1,r-1} & \mu_{m,r-1} & \mu_{m+1,r-1} & \cdots
& \mu_{r-1,r-1} & \\
\vdots & \vdots &\vdots &\vdots & \vdots & \qdots & & \\
\mu_{1,m+1} & \cdots & \mu_{m-1,m+1} & \mu_{m,m+1} & \mu_{m+1,m+1} & & & \\
\mu_{1m} & \cdots & \mu_{m-1,m} & \mu_{mm} & & & & \\
\mu_{1,m-1} & \cdots & \mu_{m-1,m-1} & & & & & \\
\vdots & \qdots & & & & & & \\
\mu_{11} & & & & & & &
\end{array}
\right)
= \left| \begin{array}{l} [\mu]^r \\[2mm] |\mu)^{r-1} \end{array} \right),
\label{mn}
\end{equation}
which satisfy the conditions
\begin{equation}
 \begin{array}{rl}
1. & \mu_{ir}\in{\mathbb Z}_+ \; \hbox{are fixed and }  \mu_{jr}-\mu_{j+1,r}\in{\mathbb Z}_+ , \;j\neq m,\;
    1\leq j\leq r-1,\\
    & \mu_{mr}\geq \# \{i:\mu_{ir}>0,\; m+1\leq i \leq r\};\\
2.& \mu_{ip}-\mu_{i,p-1}\equiv\theta_{i,p-1}\in\{0,1\},\quad 1\leq i\leq m;\;
    m+1\leq p\leq r;\\
3.  &   \mu_{mp}\geq \# \{i:\mu_{ip}>0,\; m+1\leq i \leq p\}, \quad  m+1\leq p\leq r ;\\   
4.& \hbox{if  }\;
\mu_{m,m+1}=0, \hbox{then}\; \theta_{mm}=0;  \\
5.& \mu_{ip}-\mu_{i+1,p}\in{\mathbb Z}_+,\quad 1\leq i\leq m-1;\;
    m+1\leq p\leq r-1;\\
6.& \mu_{i,j+1}-\mu_{ij}\in{\mathbb Z}_+\hbox{ and }\mu_{i,j}-\mu_{i+1,j+1}\in{\mathbb Z}_+,\\
  &  1\leq i\leq j\leq m-1\hbox{ or } m+1\leq i\leq j\leq r-1.
 \end{array}
\label{cond3}
\end{equation}

For  $m=n=1$, the new basis vectors are  given by 
\begin{equation}
|\mu) = 
 \left| \begin{array}{l} \mu_{12}, \mu_{22} \\ \mu_{11} \end{array} \right), 
\label{m2}
\end{equation}
where  
\begin{equation}
\mu_{12}\in \N \; {\rm if}\; \mu_{22}=0; \mu_{12}\in \Z_+ \; {\rm if} \; \mu_{22}\in \Z_+ \;{\rm and} \; 
\mu_{11}=\mu_{12}, \mu_{12}-1 \;({\rm if}\;  \mu_{12}=0, {\rm then} \; \mu_{11}=0) \label{cond}.
\end{equation}

We assume that the action of the $\u(1|1)$ generators is as given in~\cite{CGC} and since the weight of $|0\rangle$ is $(-\frac{p}{2} | \frac{p}{2})$, the weight
of the  vector $|\mu)$ is determined by
\[
(-\frac{p}{2}|\frac{p}{2}) + (\mu_{11}|\mu_{12}+\mu_{22}-\mu_{11}).
\]

Our aim is to compute the action of $c_1^\pm$ and $c_2^\pm$
on the basis vectors $|\mu)$. 
{}From the triple relations~\eqref{paraosp}, it follows that
under the $\u(1|1)$ basis~\eqref{un}, the set $(c_1^+, c_2^+)$ forms a tensor of rank (1,0).
 Therefore one can attach a unique GZ pattern with top line
$1 0$ to $c_1^+ $ and $c_2^+$, corresponding to the weight $\epsilon_1$ and $\delta_1$, respectevely. Explicitly  
\begin{equation}
c_1^+ \sim \left| \begin{array}{cc}1&0\\1& \end{array}\right),\qquad
c_2^+ \sim \left| \begin{array}{cc}1&0\\0& \end{array}\right).
\label{b1b2}
\end{equation}
In general for the action of $c_j^+, j=1,2$ 
on the basis vectors $|\mu)$ one can write
\begin{equation}
c_j^+| \mu )= \sum_{|\mu' )}
 ( \mu' | c_j^+ | \mu )|\mu' ), 
\end{equation}
where the matrix elements can be written as follows:
\begin{equation}
 ( \mu' | c_1^+ | \mu ) = 
\left( \begin{array}{ll} \mu_{12} & \mu_{22} \\ \mu_{11} & \end{array}; 
\begin{array}{cc}1&0\\1& \end{array} \right| \left.
\begin{array}{ll} \mu_{12}' & \mu_{22}' \\ \mu_{11}' & \end{array} \right)
\times
( \mu_{12}',\mu_{22}' || c_1^+ || \mu_{12},\mu_{22} ),
\label{matrixel1}
\end{equation}
\begin{equation}
 ( \mu' | c_2^+ | \mu ) = 
\left( \begin{array}{ll} \mu_{12} & \mu_{22} \\ \mu_{11} & \end{array}; 
\begin{array}{cc}1&0\\0& \end{array} \right| \left.
\begin{array}{ll} \mu_{12}' & \mu_{22}' \\ \mu_{11}' & \end{array} \right)
\times
( \mu_{12}',\mu_{22}' || c_2^+ || \mu_{12},\mu_{22} ).
\label{matrixel2}
\end{equation}
The first factor in the right hand side of~(\ref{matrixel1})-(\ref{matrixel2}) is a  $\u(1|1)$ Clebsch-Gordan coefficient (CGC)~\cite{CGC},
and the second factor is a {\em reduced matrix element}.
The possible values of the patterns $\mu'$ are determined by the $\u(1|1)$ tensor product
$ (1,0)\otimes (\mu_{12},\mu_{22}) = (\mu_{12}+1,\mu_{22}) \oplus (\mu_{12},\mu_{22}+1)$, and by the
additivity property of the internal labels ($\mu'_{11}=\mu_{11}+1$ in the above expression).
The only $\gl(1|1)$ CGCs of relevance are given below, their values taken from~\cite{CGC}:
\begin{align}
&\left( \begin{array}{l} \mu_{12},\mu_{22} \\ \mu_{12} \end{array}; 
 \begin{array}{l}1, 0\\1 \end{array} \right| \left.
 \begin{array}{l} \mu_{12}+1 , \mu_{22} \\ \mu_{12}+1 \end{array} \right) =
 1, \\
&\left( \begin{array}{l} \mu_{12},\mu_{22} \\ \mu_{12}-1 \end{array}; 
 \begin{array}{l}1, 0\\1 \end{array} \right| \left.
 \begin{array}{l} \mu_{12}+1, \mu_{22} \\ \mu_{12} \end{array} \right) =
 \sqrt{\frac{\mu_{12}+\mu_{22}}{\mu_{12}+\mu_{22}+1}},\\
&\left( \begin{array}{l} \mu_{12},\mu_{22} \\ \mu_{12} \end{array}; 
 \begin{array}{l}1, 0\\0 \end{array} \right| \left.
 \begin{array}{l} \mu_{12}+1 , \mu_{22} \\ \mu_{12} \end{array} \right) =
 \sqrt{\frac{1}{\mu_{12}+\mu_{22}+1}},\\
&\left( \begin{array}{l} \mu_{12},\mu_{22} \\ \mu_{12}-1 \end{array}; 
 \begin{array}{l}1, 0\\1 \end{array} \right| \left.
 \begin{array}{l} \mu_{12} , \mu_{22}+1 \\ \mu_{12} \end{array} \right) =
 -\sqrt{\frac{1}{\mu_{12}+\mu_{22}+1}},\\
&\left( \begin{array}{l} \mu_{12},\mu_{22} \\ \mu_{12} \end{array}; 
 \begin{array}{l}1, 0\\0 \end{array} \right| \left.
 \begin{array}{l} \mu_{12}, \mu_{22}+1 \\ \mu_{12} \end{array} \right) =
 \sqrt{\frac{\mu_{12}+\mu_{22}}{\mu_{12}+\mu_{22}+1}},\\ 
&\left( \begin{array}{l} \mu_{12},\mu_{22} \\ \mu_{12}-1 \end{array}; 
 \begin{array}{l}1, 0\\0 \end{array} \right| \left.
 \begin{array}{l} \mu_{12} , \mu_{22}+1 \\ \mu_{12}-1 \end{array} \right) =
 1.
\end{align}

Thus now the problem is  finding explicit expressions for the functions 
$\tilde{G}_i$ and $G_i, \;\; i=1,2$, where
\begin{align}
\tilde{G}_1(\mu)=( \mu_{12}+1,\mu_{22} || c^+_1 || \mu_{12},\mu_{22} ), \qquad
\tilde{G}_2(\mu)=( \mu_{12},\mu_{22}+1 || c^+_1 || \mu_{12},\mu_{22} ),\nn\\
G_1(\mu)=( \mu_{12}+1,\mu_{22} || c^+_2 || \mu_{12},\mu_{22} ), \qquad
G_2(\mu)=( \mu_{12},\mu_{22}+1 || c^+_2 || \mu_{12},\mu_{22} ).
\label{F1F2}
\end{align}
We can write:
\begin{align}
c^+_1 \left| \begin{array}{l} \mu_{12}, \mu_{22} \\ \mu_{12} \end{array} \right) 
& = \tilde{G}_1(\mu) 
\left| \begin{array}{l} \mu_{12}+1, \mu_{22} \\ \mu_{12}+1 \end{array} \right)
 ,\label{c11+}\\
c^+_1 \left| \begin{array}{l} \mu_{12}, \mu_{22} \\ \mu_{12}-1 \end{array} \right) 
& = \sqrt{\frac{\mu_{12}+\mu_{22}}{\mu_{12}+\mu_{22}+1}} \tilde{G}_1(\mu) 
\left| \begin{array}{l} \mu_{12}+1, \mu_{22} \\ \mu_{12} \end{array} \right)
 -\sqrt{\frac{1}{\mu_{12}+\mu_{22}+1}} \tilde{G}_2(\mu) 
\left| \begin{array}{l} \mu_{12},\mu_{22}+1 \\ \mu_{12} \end{array} \right),\label{c12+}\\
c^+_2 \left| \begin{array}{l} \mu_{12}, \mu_{22} \\ \mu_{12} \end{array} \right) 
& = \sqrt{\frac{1}{\mu_{12}+\mu_{22}+1}} G_1(\mu) 
\left| \begin{array}{l} \mu_{12}+1, \mu_{22} \\ \mu_{12} \end{array} \right)
 +\sqrt{\frac{\mu_{12}+\mu_{22}}{\mu_{12}+\mu_{22}+1}} G_2(\mu) 
\left| \begin{array}{l} \mu_{12},\mu_{22}+1 \\ \mu_{12} \end{array} \right),\label{c21+}\\
c^+_2 \left| \begin{array}{l} \mu_{12}, \mu_{22} \\ \mu_{12}-1 \end{array} \right) 
& = G_2(\mu) 
\left| \begin{array}{l} \mu_{12}, \mu_{22}+1 \\ \mu_{12}-1 \end{array} \right)
 ,\label{c22+}
\end{align}
and the action of $c_j^-$ follows from $(\mu'|c_j^-|\mu) = (\mu|c_j^+|\mu')$.

Now it remains to determine the functions $G_i$ and $\tilde{G}_i, \;\; i=1,2$.
From the action 
\begin{equation}
\{ c_2^-, c_2^+ \} |\mu) = 2h_2 |\mu) = 
(p+2(\mu_{12}+\mu_{22}-\mu_{11})) |\mu), \label{c2c2}
\end{equation}
one deduces the following recurrence relations 
for $G_1$ and $G_2$:
\begin{align}
& \frac{G_1(\mu_{12},\mu_{22}) G_2(\mu_{12}+1,\mu_{22}-1)}{\sqrt{\mu_{12}+\mu_{22}+1}}
+ \frac{G_1(\mu_{12},\mu_{22}-1) G_2(\mu_{12},\mu_{22}-1)\sqrt{\mu_{12}+\mu_{22}-1}}{\mu_{12}+\mu_{22}} =0, \label{R1}\\
& \frac{G_1(\mu_{12},\mu_{22})^2}{\mu_{12}+\mu_{22}+1}+ \frac{\mu_{12}+\mu_{22}-1}{\mu_{12}+\mu_{22}}G_2(\mu_{12},\mu_{22}-1)^2 
+\frac{\mu_{12}+\mu_{22}}{\mu_{12}+\mu_{22}+1}G_2(\mu_{12},\mu_{22})^2 =
p+2\mu_{22}, \label{R2}\\
& \frac{G_1(\mu_{12}-1,\mu_{22})^2}{\mu_{12}+\mu_{22}}+ G_2(\mu_{12},\mu_{22}-1)^2 
+G_2(\mu_{12},\mu_{22})^2 =
p+2\mu_{22}+2. \label{R3}
\end{align}
The action $c_2^-|\mu)$ leads to the boundary condition: $G_2(\mu_{12},\mu_{22}-1)=0$ if $\mu_{22}=0$.
The boundary condition together with the recurrence relations~\eqref{R1}-\eqref{R3}
lead to the following solution 
for the unknown functions $G_1$ and $G_2$:
\begin{align}
G_1(\mu_{12},\mu_{22}) & = \sqrt{ 
\frac{\mu_{12}(\mu_{12}+\mu_{22}+1)(p-\mu_{12})}{\mu_{12}+\mu_{22}}}, \quad {\rm if }\;\; \mu_{22} \; {\rm is \;\; even} \label{F1even}\\
G_1(\mu_{12},\mu_{22}) & = -\sqrt{ 
\mu_{12}(p-\mu_{12})}, \quad {\rm if }\;\; \mu_{22} \; {\rm is \;\; odd} \label{F1odd}\\
G_2(\mu_{12},\mu_{22}) & = \sqrt{ 
\mu_{12}+\mu_{22}+1}, \quad {\rm if }\;\; \mu_{22} \; {\rm is \;\; even} \label{F2even}\\
G_2(\mu_{12},\mu_{22}) & = -\sqrt{ 
\frac{(\mu_{22}+1)(p+\mu_{22}+1)}{\mu_{12}+\mu_{22}}}, \quad {\rm if }\;\; \mu_{22} \; {\rm is \;\; odd.} \label{F2odd}
\end{align}

From the action 
\begin{equation}
[ c_1^-, c_1^+ ] |\mu) = -2h_1 |\mu) = 
(p-2\mu_{12}) |\mu), \label{c1c1}
\end{equation}
one deduces the following recurrence relations 
for $\tilde{G}_1$ and $\tilde{G}_2$:
\begin{align}
& \frac{\tilde{G}_1(\mu_{12},\mu_{22}) \tilde{G}_2(\mu_{12}+1,\mu_{22}-1)}{\sqrt{\mu_{12}+\mu_{22}+1}}
- \frac{\tilde{G}_1(\mu_{12},\mu_{22}-1) \tilde{G}_2(\mu_{12},\mu_{22}-1)\sqrt{\mu_{12}+\mu_{22}-1}}{\mu_{12}+\mu_{22}} =0, \label{RR1}\\
& \frac{\mu_{12}+\mu_{22}}{\mu_{12}+\mu_{22}+1}\tilde{G}_1(\mu_{12},\mu_{22})^2- \frac{\mu_{12}+\mu_{22}-1}{\mu_{12}+\mu_{22}}\tilde{G}_1(\mu_{12}-1,\mu_{22})^2 
+\frac{\tilde{G}_2(\mu_{12},\mu_{22})^2}{\mu_{12}+\mu_{22}+1} =
p-2\mu_{12}+2, \label{RR2}\\
& \tilde{G}_1(\mu_{12},\mu_{22})^2- \tilde{G}_1(\mu_{12}-1,\mu_{22})^2-\frac{\tilde{G}_2(\mu_{12},\mu_{22}-1)^2}{\mu_{12}+\mu_{22}} 
 =
p-2\mu_{12}, \label{RR3}
\end{align}

The action $c_1^-|\mu)$ leads to the boundary condition: $G_2(\mu_{12},\mu_{22}-1)=0$ if $\mu_{22}=0$.
The boundary condition together with the recurrence relations~\eqref{RR1}-\eqref{RR3}
lead to the following solution 
for the unknown functions $\tilde{G}_1$ and $\tilde{G}_2$:
\begin{align}
\tilde{G}_i(\mu_{12},\mu_{22})=G_i(\mu_{12},\mu_{22}), \;\; i=1,2, \quad {\rm if }\;\; \mu_{22} \; {\rm is \;\; even}, \label{FF1even}\\
\tilde{G}_i(\mu_{12},\mu_{22})=-G_i(\mu_{12},\mu_{22}), \;\; i=1,2, \quad {\rm if }\;\; \mu_{22} \; {\rm is \;\; odd}. \label{FF2even}
\end{align}

The solution for $\tilde{G}_i$ and $G_i, \;\; i=1,2$ is unique up to a choice of the sign factor. 
At this moment, only the actions of $[c_1^-,c_1^+]$ and $\{c_2^-,c_2^+\}$ have been used in the process.
Now it remains to verify whether the actions of $c_1^\pm$ and $c_2^\pm$ thus determined
do indeed yield a solution, i.e.\ one should verify that all triple relations 
are satisfied. This is a straightforward but tedious computation; the only result provided
by this calculation is that the sign factors are restricted and their choice 
in~(\ref{F1even})-(\ref{F2odd}), (\ref{FF1even})-(\ref{FF2even}) is the simplest solution.

The explicit expressions for the reduced matrix elements~(\ref{F1even})-(\ref{F2odd}) 
and~(\ref{FF1even})-(\ref{FF2even}) give the
action of the generators in the basis of $\overline V(p)$, for arbitrary $p$.
The structure of the maximal submodule $M(p)$ and hence of the irreducible
factor module $V(p)$ is revealed by examining when these matrix elements vanish.
The only crucial factor is
\[
(p-\mu_{12}).
\]
So,  starting from the vacuum vector, with a GZ-pattern 
consisting of all zeros, one can raise the entries in the GZ-pattern by applying
the operators $c_j^+$. However, when $\mu_{12}$ has reached the value~$p$ it can no
longer be increased. As a consequence, all vectors $|\mu)$ with $\mu_{12}>p$ belong to
the submodule $M(p)$. This uncovers the structure of $V(p)$.
We summarize:

\begin{theo}
\label{prop-main}
An orthonormal basis for the parastatistics Fock space $V(p)$ of one pair of parafermions and one pair of 
parabosons is given by the vectors $|\mu)$,
see~(\ref{m2})-(\ref{cond}), with $\mu_{12}\leq p$. 
The action of the Cartan algebra elements of $\osp(3|2)$ is:
\begin{equation}
h_{1}|\mu)=\left(-\frac{p}{2}+\mu_{11}\right)|\mu), 
\quad h_{2}|\mu)=\left(\frac{p}{2}+\mu_{12}+\mu_{22}-\mu_{11}\right)|\mu). \label{h_k} \\
\end{equation}
For the action of the parastatistics operators $c_j^+, \; j=1,2$ we have:
\begin{align}
c^+_1 \left| \begin{array}{l} \mu_{12}, \mu_{22} \\ \mu_{12} \end{array} \right) 
& = (-1)^{\mu_{22}}{G}_1(\mu_{12},\mu_{22}) 
\left| \begin{array}{l} \mu_{12}+1, \mu_{22} \\ \mu_{12}+1 \end{array} \right)
 ,\label{2c11+}\\
c^+_1 \left| \begin{array}{l} \mu_{12}, \mu_{22} \\ \mu_{12}-1 \end{array} \right) 
& = \sqrt{\frac{\mu_{12}+\mu_{22}}{\mu_{12}+\mu_{22}+1}} (-1)^{\mu_{22}}{G}_1(\mu_{12},\mu_{22}) 
\left| \begin{array}{l} \mu_{12}+1, \mu_{22} \\ \mu_{12} \end{array} \right) \nn\\
 &-\sqrt{\frac{1}{\mu_{12}+\mu_{22}+1}} (-1)^{\mu_{22}}{G}_2(\mu_{12},\mu_{22}) 
\left| \begin{array}{l} \mu_{12},\mu_{22}+1 \\ \mu_{12} \end{array} \right),\label{2c12+}\\
c^+_2 \left| \begin{array}{l} \mu_{12}, \mu_{22} \\ \mu_{12} \end{array} \right) 
& = \sqrt{\frac{1}{\mu_{12}+\mu_{22}+1}} G_1(\mu_{12},\mu_{22}) 
\left| \begin{array}{l} \mu_{12}+1, \mu_{22} \\ \mu_{12} \end{array} \right)\nn\\
 &+\sqrt{\frac{\mu_{12}+\mu_{22}}{\mu_{12}+\mu_{22}+1}} G_2(\mu_{12},\mu_{22}) 
\left| \begin{array}{l} \mu_{12},\mu_{22}+1 \\ \mu_{12} \end{array} \right),\label{2c21+}\\
c^+_2 \left| \begin{array}{l} \mu_{12}, \mu_{22} \\ \mu_{12}-1 \end{array} \right) 
& = G_2(\mu_{12},\mu_{22}) 
\left| \begin{array}{l} \mu_{12}, \mu_{22}+1 \\ \mu_{12}-1 \end{array} \right)
 ,\label{2c22+}
\end{align}
and the action of the parastatistics operators $c_j^-, \; j=1,2$ is given by:
\begin{align}
c^-_1 \left| \begin{array}{l} \mu_{12}, \mu_{22} \\ \mu_{12} \end{array} \right) 
& = (-1)^{\mu_{22}}{G}_1(\mu_{12}-1, \mu_{22}) 
\left| \begin{array}{l} \mu_{12}-1, \mu_{22} \\ \mu_{12}-1 \end{array} \right)\nn\\
&+\sqrt{\frac{1}{\mu_{12}+\mu_{22}}} (-1)^{\mu_{22}}{G}_2(\mu_{12}, \mu_{22}-1) 
\left| \begin{array}{l} \mu_{12},\mu_{22}-1 \\ \mu_{12}-1 \end{array} \right) ,\label{2c11-}\\
c^-_1 \left| \begin{array}{l} \mu_{12}, \mu_{22} \\ \mu_{12}-1 \end{array} \right) 
& = \sqrt{\frac{\mu_{12}+\mu_{22}-1}{\mu_{12}+\mu_{22}}} (-1)^{\mu_{22}}{G}_1(\mu_{12}-1, \mu_{22}) 
\left| \begin{array}{l} \mu_{12}-1, \mu_{22} \\ \mu_{12}-2 \end{array} \right) 
 ,\label{2c12-}\\
c^-_2 \left| \begin{array}{l} \mu_{12}, \mu_{22} \\ \mu_{12} \end{array} \right) 
& = \sqrt{\frac{\mu_{12}+\mu_{22}-1}{\mu_{12}+\mu_{22}}} G_2(\mu_{12}, \mu_{22}-1) 
\left| \begin{array}{l} \mu_{12},\mu_{22}-1 \\ \mu_{12} \end{array} \right),\label{2c21-}\\
c^-_2 \left| \begin{array}{l} \mu_{12}, \mu_{22} \\ \mu_{12}-1 \end{array} \right) 
&=\sqrt{\frac{1}{\mu_{12}+\mu_{22}}} G_1(\mu_{12}-1, \mu_{22}) 
\left| \begin{array}{l} \mu_{12}-1, \mu_{22} \\ \mu_{12}-1 \end{array} \right)\nn\\
& + G_2(\mu_{12},\mu_{22}-1) 
\left| \begin{array}{l} \mu_{12}, \mu_{22}-1 \\ \mu_{12}-1 \end{array} \right)
 ,\label{2c22-}
\end{align}
where $G_i,\ i=1,2,$ are given by Eqs.~(\ref{F1even})-(\ref{F2odd}).

\end{theo}

\setcounter{equation}{0}
\section{Summary and conclusion} \label{sec:summary}

In this paper we have investigated the Fock spaces $V(p)$ of $m$ parafermions 
and $n$ parabosons  with relative parafermion relations among them,   which 
are  the unitary irreducible
representations  of $\osp(2m+1|2n)$ with lowest weight 
$(-\frac{p}{2},\ldots,-\frac{p}{2}|\frac{p}{2},\ldots,\frac{p}{2})$.
We have used group theoretical methods and computational techniques.
A crucial role in the analysis is played by the $\u(m|n)$
subalgebra of $\osp(2m+1|2n)$, generated by all supercommutators of the parafermions
and parabosons. Taking a certain parabolic subalgebra
${\cal P}$ containing $\u(m|n)$ and a trivial module of ${\cal P}$ generated from
the vacuum, i.e.\ the lowest weight vector of weight $(-\frac{p}{2},\ldots,-\frac{p}{2}|\frac{p}{2},\ldots,\frac{p}{2})$ ,
an induced module $\overline V(p)$ of $\osp(2m+1|2n)$ is constructed. 
The Fock space $V(p)$ is the
quotient of this induced module by its maximal submodule $M(p)$. The
character of the induced module is  obtained and is rewritten as an
infinite sum over certain partitions of supersymmetric Schur functions. This can be
reinterpreted as a decomposition of the $\osp(2m+1|2n)$ module into an infinite
sum of finite-dimensional simple $\u(m|n)$ modules labeled  by partitions, namely the  covariant $\u(m|n)$
tensor modules. For
each such representation of $\u(m|n)$ one can use the corresponding
Gelfand-Zetlin basis. The union of all these GZ basis vectors is the basis for
the induced module $\overline V(p)$. The main calculation is then the action
of one pair of parafermion and one pair of paraboson operators on this basis.
In order to  calculate the matrix elements, they are written as a product
of certain $\u(1|1)$ Clebsch-Gordan coefficient and a reduced
matrix element. As the relevant $\u(1|1)$ (in general $\u(m|n)$) CGCs are known, the problem is to find
the reduced matrix elements. Solving a set of recurrence relations for these, leads
to their expressions. The last give not only the
action of the generators in the basis of $\overline V(p)$, they also yield
the structure of the maximal submodule $M(p)$ and hence of the irreducible factor module
$V(p)$. This leads to  an explicit basis of $V(p)$ (consisting
of all possible GZ-patterns with $\mu_{12}$ integer  at most $p$) and the explicit
action of the generators in this basis. 

We have found the matrix elemenets only in the case $m=n=1$.
The real interest lies in such quantum systems (parabosons and parafermions)
with any degree of freedom, including  infinite degree of freedom.
We hope to be able to find the matrix elements for any $m$ and $n$ and to report the result soon.

\section*{Acknowledgments}
The author would like to thank Professor J. Van der Jeugt, Professor R.C. King and Professor T.D.~Palev 
for constructive discussions and their interest.

\end{document}